# Local motifs in GeS$_2$-Ga$_2$S$_3$ glasses

I. Pethes[a,*], V. Nazabal[b], R. Chahal[b], B. Bureau[b], I. Kaban[c], S. Belin[d], P. Jóvári[a]

[a]Wigner Research Centre for Physics, Hungarian Academy of Sciences, H-1525 Budapest, POB 49, Hungary

[b]Institut Sciences Chimiques de Rennes, UMR-CNRS 6226, Campus de Beaulieu, Université de Rennes 1, 35042 Rennes Cedex, France

[c]IFW Dresden, Institute for Complex Materials, Helmholtzstr. 20, 01069 Dresden, Germany

[d]Synchrotron SOLEIL, L'Orme des Merisiers, Saint Aubin, 91192 Gif sur Yvette, France

Abstract

The structure of (GeS$_2$)$_{0.75}$(Ga$_2$S$_3$)$_{0.25}$ and (GeS$_2$)$_{0.83}$(Ga$_2$S$_3$)$_{0.17}$ glasses was investigated by Raman scattering, high energy X-ray diffraction and extended X-ray absorption fine structure (EXAFS) measurements at the Ga and Ge K-edges. The reverse Monte Carlo simulation technique (RMC) was used to obtain structural models compatible with diffraction and EXAFS datasets. It was found that the coordination number of Ga is close to four. While Ge atoms have only S neighbors, Ga binds to S as well as to Ga atoms showing a violation of chemical ordering in GeS$_2$-Ga$_2$S$_3$ glasses. Analysis of the corner- and edge-sharing between [GeS$_{4/2}$] units revealed that about 30% of germanium atoms participate in the edge-shared tetrahedra.

## 1. Introduction

Sulfide glasses are pivotal for optical applications because of their unique properties such as a large transparency window in the visible and mid-infrared range extending to the wavelengths far beyond those of oxide glasses, a low phonon energy, high linear and nonlinear refractive indices or photosensitivity [1-4]. The passive optical properties of these glasses can be tailored by their chemical compositions, and applicability of gallium-based glasses can be further enhanced by rare earth doping into the glassy network. As active optical media, they could be used as source of radiation in different utilizations such as mid-IR fiber lasers, optical amplifiers and upconverters [5-8].

It has been found that glasses containing gallium can dissolve relatively larger amounts of rare earth elements than other chalcogenide glasses [9-11]. This phenomenon seems to be related to their local electronic and molecular structure induced by the addition of Ga [12]. Therefore, it is crucial for further applications to understand their structural properties properly.

---

[*] Corresponding author. E-mail address: pethes.ildiko@wigner.mta.hu

The mean coordination number, which is the sum of the coordination numbers of the constituents weighted by their concentrations, has a significant impact on the physical and chemical properties of a glass [13-15]. In glasses consisting of elements of group 14, 15 and 16 the coordination numbers of the components follow the 8-N Mott rule [16], where N is the formal chemical valence (e.g. Ge-As-Se [17], Ge-As-Te [18], Ge-Sb-Te [19]). However, glasses containing a group 13 component can deviate from this rule as the coordination numbers of Ga or In can be four or even higher instead of threefold coordination predicted by the Mott-rule ([20- 22]).

Fourfold coordinated Ga atoms were also observed in various crystalline gallium chalcogenides (e.g. [$GaS_4$] tetrahedra in crystalline $Ga_2S_3$ [23], Ca(Sr,Ba)$Ga_2S_4$ [24], [$GaSe_4$] tetrahedra in $Ga_2Se_3$ [25], and [Se(S)$_3$Ga-GaSe(S)$_3$] units in GaS and GaSe [26]). These investigations also indicated that a certain number of the chalcogen atoms can be three or fourfold coordinated [24, 26].

Another characteristic property of a glassy network is its chemical order: while some systems behave as a random covalent network without clear preferential bonding (e.g. Ge-As-Se [27]), others can be better described with the chemically ordered network model [28], in which bonds with higher bond energy are preferred (e.g. Ge-As-S [29]). According to the latter, chalcogen atoms bind preferentially to metal (Ga) or metalloid (Ge, As) atoms. (Metal and metalloid atoms will be denoted with M throughout the paper.) Considerable amount of chalcogen-chalcogen or M-M bonds exist only in chalcogen rich or chalcogen poor compositions, respectively.

Ge-Ga-Ch (Ch=S, Se or Te) glasses were the focus of several studies in recent years [12, 21, 30-44]. These investigations mostly agree that the coordination numbers of Ge and Ga are four. The propensity of forming various bonds was also investigated to verify the validity of chemically ordered network model. According to that model, [$GeCh_4$] and [$GaCh_4$] tetrahedral units are connected together by corner-sharing or edge-sharing in stoichiometric compounds, supplemented by chalcogen chains or rings in chalcogen-rich systems or by [$Ch_3Ge(Ga)-(Ga)GeCh_3$] ethane-like units in chalcogen-deficient systems. It should be noted that if Ga is fourfold coordinated then compositions along the $(GeCh_2)_x(Ga_2Ch_3)_{(1-x)}$ tie line are in fact in deficit of chalcogen.

In Te-rich Ga-Ge-Te systems, Raman spectroscopy, X-ray diffraction (XRD), neutron diffraction (ND) and extended X-ray absorption fine structure (EXAFS) measurements combined with simulations have shown tetrahedral arrangement around Ge and Ga, and Te coordination number significantly higher than two [30]. They found only negligible amount of M-M bonds in these chalcogen-rich compounds.

Ge-Ga-Se systems were investigated by several groups and methods, for example by Raman scattering [31-34], X-ray diffraction and IR spectroscopy [35], X-ray photoelectron spectroscopy (XPS) [32, 33], EXAFS [33], NMR spectroscopy [34], neutron diffraction and EXAFS measurement combined with reverse Monte Carlo (RMC) simulation technique [21]. Ge and Ga atoms were found to be fourfold coordinated [21, 31, 33-35]. The average coordination number of selenium atoms was found to be higher than two in several studies [21, 32, 34], but twofold coordinated Se-atoms were reported by Golovchak et al. [33]. The chemically ordered network model seems to be more or less valid in these compounds as reported in Refs. [31, 35, 33], but significant amount of M-M bonds were found even in Se-rich systems by Pethes et al. [21]. The Ge-Ge, Ge-Ga and Ga-Ga bonds are hardly distinguishable by Raman measurements, but EXAFS measurements combined with diffraction investigations and RMC suggest that Ge-Ga bonds are preferred among the M-M bonds in these glasses.

There are a lot of experimental studies devoted to the structure of Ge-Ga-S glasses. These works also use several different methods such as Raman [12, 36- 39], XAFS [40], neutron diffraction [41], EXAFS [42], XRD and Raman [43], 2D micro-Raman [44]. They have all found that Ge and Ga atoms are equally fourfold coordinated, $[GaS_4]$ and $[GeS_4]$ tetrahedral units are cross-linked via bridging sulfur atoms, M-chalcogen bonds are preferable. $[S_3(Ga/Ge)-(Ga/Ge)S_3]$ ethane-like units were observed to form as a consequence of the sulfur deficiency, but the type of the M-M bonds is uncertain. The latter can either be Ge-Ge [41], Ga-Ga [43], Ge-Ga or Ga-Ga [44], but they are found to be indistinguishable by most papers. Simultaneous existence of a small amount of M-M and S-S bonds was reported indicating some chemical disorder [37, 39, 41]. Triply coordinated S atoms were found by Masselin et al. [44].

In this work we investigate $(GeS_2)_{0.75}(Ga_2S_3)_{0.25}$ and $(GeS_2)_{0.83}(Ga_2S_3)_{0.17}$ glasses using Raman-spectroscopy, X-ray diffraction and EXAFS measurements. Large scale structural models are obtained by fitting EXAFS and diffraction data sets simultaneously with the RMC simulation technique [45, 46]. Bond preferences (especially the type of M-M bonds), coordination numbers and nearest neighbor distances for Ge, Ga and S atoms, formation of the corner- and edge-sharing tetrahedra are analyzed in detail.

2. Experimental

## 2.1 Sample preparation

For synthesis of $(GeS_2)_{0.75}(Ga_2S_3)_{0.25}$ and $(GeS_2)_{0.83}(Ga_2S_3)_{0.17}$ glasses, high purity (5N) germanium, gallium and sulfur were used. Despite of the purity of the commercial material, sulfur can be polluted by water and carbon. Therefore, sulfur was additionally purified by distillation. Then, the proper quantities of the chemical reagents were placed in silica tubes and sealed under vacuum. The batch was slowly heated and homogenized in a rocking furnace for 8 h at 950°C. Glass rods were obtained by cooling the silica tubes with alloys in water. They were then annealed near the glass transition temperature for 2 hours. Several glass discs of about 1 mm thickness and 15 mm in diameter were cut from the annealed rods. The composition of each sample was analyzed by means of scanning electron microscopy with an energy-dispersive X-ray analyzer (Oxford Instruments). The obtained values agreed with the nominal composition within ±0.5 at.%.

## 2.2 Measurements

The investigated glass compositions and their densities are given in Table 1. Density was determined by means of a Mettler Toledo XS64 system measuring the weight of the samples in air and water. The accuracy of the density values is ±0.005 g/cm$^3$.

| composition | density (g/cm$^3$) | atomic density (at/Å$^3$) |
|---|---|---|
| $(GeS_2)_{0.75}(Ga_2S_3)_{0.25}$ | 2.955 | 0.03857 |
| $(GeS_2)_{0.83}(Ga_2S_3)_{0.17}$ | 2.891 | 0.03787 |

**Table 1** Investigated compositions and their densities.

X-ray diffraction (XRD) measurements were carried out at the BW5 wiggler beamline [47] of HASYLAB, DESY (Hamburg, Germany). The energy of monochromatic synchrotron radiation was 85 keV ($\lambda$ = 0.146 Å). Powder samples were placed into thin walled (20 μm) quartz capillaries with outer diameter of 2 mm. The cross section of the incident beam was 1×2 mm$^2$. Scattered intensities were recorded by a Ge solid state detector. Raw data were corrected for background, polarization, absorption and Compton scattering. X-ray diffraction structure factors are shown in Figs. 1 and 2.

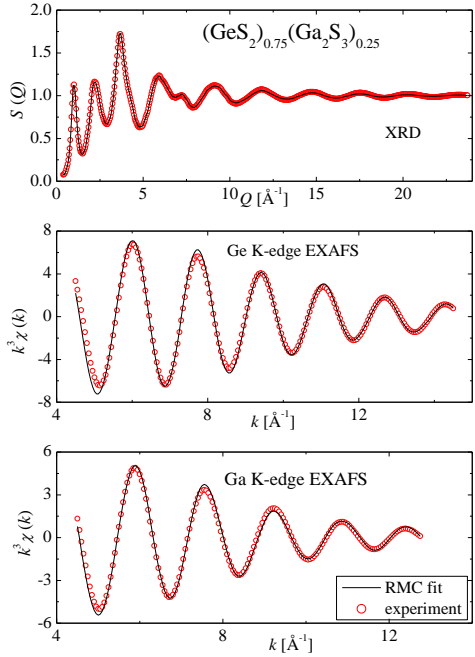
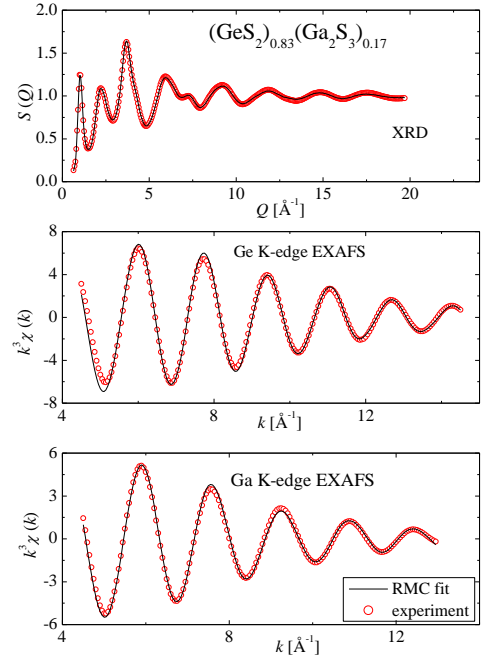

**Figure 1** X-ray diffraction structure factor, and $k^3$ weighted, filtered Ge and Ga K-edge EXAFS spectra (open symbols) and fits for the 'final' model (lines) of $(GeS_2)_{0.75}(Ga_2S_3)_{0.25}$ sample.

**Figure 2** X-ray diffraction structure factor, and $k^3$ weighted, filtered Ge and Ga K-edge EXAFS spectra (open symbols) and fits for the 'final' model (lines) of $(GeS_2)_{0.83}(Ga_2S_3)_{0.17}$ sample.

EXAFS spectra of powdered glasses were recorded at the SAMBA beamline of SOLEIL. The EXAFS measurements were performed in transmission or fluorescence modes, at room temperature for Ge K-edge and a liquid $N_2$ cryostat was used for Ga K-edge. Raw intensities were converted into $\chi(k)$ curves by the Viper program [48]. Raw $\chi(k)$ data (see Figure 1) were filtered in two steps: first $k^3\chi(k)$ was forward Fourier-transformed into r-space using a Kaiser-Bessel window (α=1.5). The *k*-range of transformation was 1.85 Å$^{-1}$ - 12.8 Å$^{-1}$ for the Ga edge and 1.85 Å$^{-1}$ - 14.5 Å$^{-1}$ for the Ge edge. The resulting *r*-space data were backtransformed using a rectangular window between 0.85 Å and 2.6 Å. Backscattering amplitudes and phases needed to calculate the model curves were obtained by the Feff program [49].

The Raman scattering data were recorded at ambient temperature on glass samples by an HR800 (Horiba–Jobin-Yvon) unpolarized confocal micro-Raman spectrophotometer with 785 nm laser diode at room temperature using low power density. Reduced Raman intensity of glasses Raman spectra were calculated considering the following equation $I_{red}(\omega) = I(\omega)\omega/[n(\omega) + 1]$ [36, 50]. The term $I(\omega)$ represents the experimental Raman intensity at $\omega$ pulsation, and $n(\omega)$ is the Bose–Einstein factor, defined as $n(\omega) = (\exp[(\hbar\omega/kT)-1])-1$, where $\hbar$ is Planck's constant.

## 3. Reverse Monte Carlo simulations

The reverse Monte Carlo simulation technique is a framework for combining the information content of various experimental datasets (e.g. neutron diffraction, X-ray diffraction and EXAFS) with prior physical and chemical knowledge (density, valence, bond angles). RMC works with a large atomic configuration (typically 10000-100000 atoms), which is modified by random atomic moves according to the following scheme:

0. Generation of starting configuration: it can either be a completely random set of points in a box, a crystalline model or the result of a previous simulation (RMC, molecular dynamics)
1. Move an atom at random and check whether the minimum interatomic distances (cut offs) are violated. Discard the move if the cut offs are not satisfied and generate a new move.
2. If the cut offs are satisfied then calculate partial pair correlation functions, model diffraction and EXAFS curves, coordination numbers, bond angle distributions or any other quantity of interest.

The $S_{mod}(Q)$ model structure factor can be obtained from the $g_{ij}(r)$ partial pair correlation functions by the following equations:

$$S_{ij}(Q) = 1 + \frac{4\pi\rho_0}{Q} \int r\sin Qr \left(g_{ij}(r)-1\right) dr \tag{1}$$

$$S_{mod}(Q) = \sum_{i \leq j} w_{ij}(Q) S_{ij}(Q) \tag{2}$$

(In case of neutron diffraction the $w_{ij}$ weighting factors do not depend on $Q$ thus summation is carried out in real space and the resulting total pair correlation function is transformed $S_{mod}(Q)$ in a way similar to Equation 1.)

For EXAFS first the $\chi_{ij}$ 'partial EXAFS curves' are calculated according to Equation 3:

$$\chi_{ij}(k) = 4\pi\rho c_j \int r^2 \gamma_{ij}(k,r) g_{ij}(r) dr \tag{3}$$

Here $i$ is the index of the absorbing component, $\varrho$ is the number density, $c_j$ is the concentration of atoms type $j$ and $\gamma_{ij}(k,r)$ is the $k$- and $r$-dependent response function of a $j$ type backscatterer. $\gamma(k,r)$ matrix can be obtained by dedicated programs such as feff [49].

The $\chi^i_{\text{mod}}(k)$ model EXAFS curve is obtained by summing the partial EXAFS curves:

$$\chi^i_{\text{mod}}(k) = \sum_j \chi_{ij}(k) \qquad (4)$$

For details of fitting EXAFS data with RMC we refer to [51].

3. The new model curves, coordination numbers etc. are compared to the experimental curves and target coordination numbers (coordination constraints). A move is always accepted if the general agreement gets better. Otherwise it is accepted with a probability between 0 and 1 [46].

4. Go back to step 1.

Due to equations 1 and 3 the pivotal quantities are the partial pair correlation functions. Unless no specific constraints are used higher order correlations such as bond angles, dihedral angles or the presence of various local motifs (e.g. 'ethane like units') cannot be subtracted reliably *from the model configuration*. For example, it has been shown that the structure factor of amorphous Si can be equally fitted with 100% and 0% tetrahedral coordination [52]. On the other hand, the average coordination number is close to 4 in the latter, obviously non-physical case as well. The reason is very clear: while the existence of Si-Si$_{4/2}$ tetrahedral units belongs to the realm of many-body correlations the average coordination number is directly related to the pair correlation function of amorphous Si.

We note that this is not a drawback of RMC: any method depending merely on diffraction or EXAFS meets this fundamental limitation. A great advantage of RMC is that by using various constraints (e.g. on coordination numbers) this limitation can be at least partly overcome.

Model configurations were generated by fitting simultaneously the Ge and Ga K-edge EXAFS datasets and the XRD structure factor for each composition by the RMC++ code [46]. The simulation boxes contained around 20000 atoms.

Dedicated runs were carried out to decide which bond types are needed to get reasonable fits. Ge-S and Ga-S bonds were allowed in every tested model. Allowing S-S bonds led to unrealistic bond lengths in the test runs; therefore S-S bonding was forbidden in further modeling. The absence of S-S bonds is also in line with the composition of the samples, which are on the GeS$_2$-Ga$_2$S$_3$ tie-line. It also follows from the compositions that if Ge is fourfold coordinated and S is twofold coordinated then the average coordination number of Ga can be higher than 3 only if M-M bonds exist. Ge-Ge, Ge-Ga and Ga-Ga bonds were allowed in different combinations to check their influence on fit quality. (A bond is forbidden if the corresponding cutoff is higher than the expected value of the bond length - e.g. 2.75 Å vs. 2.45 Å for Ge-Ge pairs.) The applied minimum interatomic distances are presented in Table 2.

| pair | Ge-Ge | Ge-Ga | Ge-S | Ga-Ga | Ga-S | S-S |
|---|---|---|---|---|---|---|
| bond allowed | 2.35 | 2.35 | 2.00 | 2.45 | 2.05 | -- |
| bond forbidden | 2.75 | 3.1 | -- | 3.1 | -- | 3.15 |

**Table 2** Minimum interatomic distances (cut offs) of the different atom pairs in Angström.

The σ parameters used to calculate the RMC cost function [45] were reduced in three steps to the final values of $5 \cdot 10^{-4}$ for XRD and of $1 \cdot 10^{-5}$ for the EXAFS data sets in every simulation runs. The number of accepted moves was around $2\text{-}3 \cdot 10^{7}$.

Some 'background' coordination constraints were always applied to avoid unrealistically high coordination numbers (such as 7 or more neighbors for Ge and Ga, 4 or more neighbors for S) or segregated atoms (atoms with zero neighbors).

The resulting models were assessed by comparing their *R*-factors to the corresponding values of a reference model. In the reference model every M-M bond was allowed and only the 'background' coordination constraints were applied. The cumulative relative *R*-factor ($R_c$), which is the average of the relative *R*-factors (*R*(investigated model)/*R*(reference model)) of the data sets [27], was used to compare the different models. The $R_c$ of the reference model is 1, by definition.

## 4. Results and discussion
### 4.1. Raman scattering spectroscopy analyses

The Raman spectra of the two Ga-Ge-S glasses (($GeS_2$)$_{0.75}$($Ga_2S_3$)$_{0.25}$ and ($GeS_2$)$_{0.83}$($Ga_2S_3$)$_{0.17}$) (see Fig. 3) can be compared in first approach to $GeS_2$ glass which has been often reported in previous studies [53- 55]. Their spectra are dominated by the presence of a broad band composed of a main peak at 342 cm$^{-1}$ related to the $v_1$ symmetric stretching modes of [$GeS_{4/2}$] tetrahedra and a shoulder observed at 370 cm$^{-1}$, whose origin remains controversial. Usually it is assigned to $v^c_1(A^c_1)$ companion mode of the $v_1(A_1)$ mode and more rarely is related to vibrations of tetrahedra sharing edge [55]. Indeed, the 370 cm$^{-1}$ and 405 cm$^{-1}$ were also attributed to splitting of molecular mode $v_3(F_2)$ due to intermolecular coupling [53]. The band at 433 cm$^{-1}$ is classically assigned to vibrations of [$S_{3/2}$Ge-S-GeS$_{3/2}$] units where the tetrahedra are connected by their corners in rings or other extended structures [36, 56].

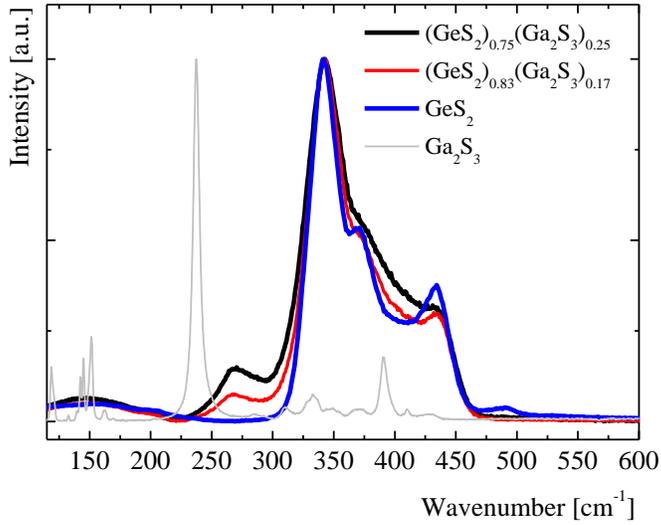

**Figure 3** Raman spectra of the crystalline α-Ga$_2$S$_3$, GeS$_2$, (GeS$_2$)$_{0.75}$(Ga$_2$S$_3$)$_{0.25}$, and (GeS$_2$)$_{0.83}$(Ga$_2$S$_3$)$_{0.17}$ glasses at 785 nm laser excitation. All curves are normalized to the intensity of the strongest band.

The relative weights of Ge and Ga suggest that two of the frequencies of normal vibrational mode of [GaS$_4$] can be expected in the vicinity of those of [GeS$_4$] tetrahedra and was proposed to be located at 350 cm$^{-1}$ $v_1(A_1)$ and 390 cm$^{-1}$ $v_3(F_2)$ or $v^c_1(A_1)$ [36]. Ishibashi et al. [57] measured low resolution spectra of RS-Ga$_2$S$_3$ glasses (R =Ca, Sr, Ba) and proposed a different assignment based on analysis of α-Ga$_2$S$_3$, SrGa$_2$S$_4$ and BaGa$_2$S$_4$ crystals. The main peak at 233 cm$^{-1}$ of α-Ga$_2$S$_3$ phase is in good agreement with our spectrum peaking at 237 cm$^{-1}$ and 390 cm$^{-1}$. In α-Ga$_2$S$_3$ the [GaS$_4$] tetrahedra are linked by corners and S atoms present two-fold and three-fold coordination and the main peak is attributed to the stretching vibration of [GaS$_4$] tetrahedra ($v_s$ [GaS$_4$](A$_1$); vibrations of the sulfur anions) and 390 cm$^{-1}$ to the deformation vibration ($v_d$ [GaS$_4$](F$_2$)) [58]. The SrGa$_2$S$_4$ crystal with [GaS$_4$] corner and edge sharing tetrahedra presents two main peaks at 280 and 357 cm$^{-1}$ related to $v_s$ vibration modes of two [GaS$_4$] tetrahedra connected by edge and $v_d$(Ga-S-Ga) of [GaS$_4$] connected by corners, respectively. The BaGa$_2$S$_4$ crystal is a 3D framework with only corner sharing [GaS$_4$] tetrahedra and the 303 cm$^{-1}$ peak is attributed to $v_d$ [GaS$_4$] (F$_2$) deformation mode. Based on their assignments for these reference crystals, they associated for RS-Ga$_2$S$_3$ glasses a band at 318 cm$^{-1}$ to the $v_d$ [GaS$_4$] (F$_2$) or $v_s$ [GaS$_4$] edge-sharing and 358-378 cm$^{-1}$ to (Ga-S-Ga) deformation mode of [GaS$_4$] corners sharing.

The lack of resolution in this range of the spectrum makes difficult precise assignments. Nonetheless, it was effectively shown that moderate substitution of $Ga_2S_3$ to $GeS_2$ glass-former leads to the restructuring of band intensities in the 340-435 $cm^{-1}$ range [37]. An obvious widening of the dominant peak at 342 $cm^{-1}$ is clearly visible for composition with highest amount of gallium, related to the presence of vibrational modes of $[GaS_{4/2}]$ tetrahedra. The shoulders at 370 $cm^{-1}$ and 433 $cm^{-1}$ both decrease in intensity, while a component around 390 $cm^{-1}$ grows with increasing $Ga_2S_3$ concentration (see Figure 3).

The two glasses studied in this work present a deficit in sulfur if we consider that Ga and Ge atoms are mainly inserted into tetrahedral sites and thus, the S-deficit is compensated by the formation of $[S_3Ge(Ga)-Ge(Ga)S_3]$ entities as proposed by previous studies [57]. The Ge-Ge vibrational mode is located at 258 $cm^{-1}$ and the shift of the lower frequency peak from 258 $cm^{-1}$ to 268 $cm^{-1}$ is related to the introduction of $Ga_2S_3$. Thus, according to Raman analysis, it can be supposed that the formation of Ga-Ga or Ga-Ge bonds is preferred over Ge-Ge bonds.

## 4.2. RMC investigations
### 4.2.1. Unconstrained simulations

The measured XRD structure factors and $k^3$ weighted $\chi(k)$ curves of the investigated compositions are shown in Figs. 1-2. Various model configurations were obtained by the simultaneous fit of the three experimental data sets for each composition. Simulations showed that the experimental data sets can be properly fitted when Ge-Ge and Ge-Ga bonds are forbidden and only Ga-Ga bonds are allowed. (Hereafter this model will be called as the 'final' model.) Model curves are presented also in Figs. 1-2. The presence of the other two M-M type bonds did not enhance the quality of the fits. The lower $R_c$ values for the models with less allowed bond types clearly indicate that the eliminated bond types (Ge-Ge, Ge-Ga, S-S bonds) are not needed to get consistent models (even if some Ge-Ge and Ge-Ga bonds may survive due to the inherently probabilistic nature of RMC).

In the unconstrained simulation runs, the coordination numbers of the atoms were free, and only the above described 'background' coordination constraints were applied. Partial pair correlation functions calculated for the 'final' model are shown in Fig. 4. The average coordination numbers obtained by the application of this model are given in Table 3. For both sulfide glasses, it was found that germanium and sulfur atoms mostly obey the 8-N rule, and their total coordination numbers are near four and two, respectively ($N_{Ge}$=3.94 and 4.07, $N_S$=2.03 and 2.15 for $(GeS_2)_{0.83}(Ga_2S_3)_{0.17}$ and $(GeS_2)_{0.75}(Ga_2S_3)_{0.25}$ compositions). The average coordination number of gallium atoms was also close to 4, though with a somewhat higher uncertainty ($N_{Ga}$=3.67 and 3.85).

|  | $N_{Ge-S}$ | $N_{S-Ge}$ | $N_{Ga-S}$ | $N_{S-Ga}$ | $N_{Ga-Ga}$ | $N_{Ge}$ | $N_{Ga}$ | $N_S$ |
|---|---|---|---|---|---|---|---|---|
| $(GeS_2)_{0.75}(Ga_2S_3)_{0.25}$ | 4.07 | 1.36 | 3.56 | 0.79 | 0.29 | 4.07 | 3.85 | 2.15 |
| $(GeS_2)_{0.83}(Ga_2S_3)_{0.17}$ | 3.94 | 1.51 | 3.32 | 0.52 | 0.35 | 3.94 | 3.67 | 2.03 |

**Table 3** $N_{ij}$ coordination numbers and $N_i$ total coordination numbers obtained by RMC for the 'unconstrained' 'final' model.

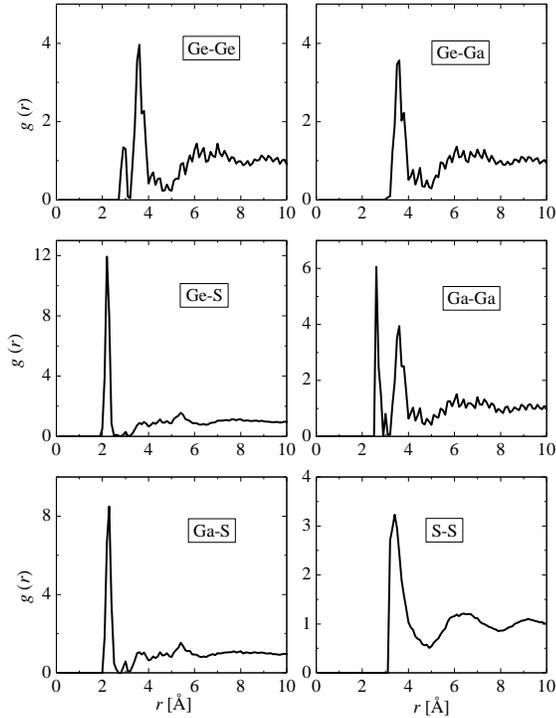

**Figure 4** Partial pair correlation functions for the 'final' model of $(GeS_2)_{0.83}(Ga_2S_3)_{0.17}$ sample.

### 4.2.2. Constrained simulations

The above results already strongly suggest that Ga is fourfold coordinated and Ga-Ga bonds exist in the compositions investigated. However, due to the similar scattering properties of Ge and Ga, changes of Ge-S and Ga-S coordination numbers may compensate each other increasing the uncertainty of Ga-related short range order parameters. This effect can be minimized by constraining the total coordination number of atoms. Therefore, in the next step, coordination constraints were used. Ge, Ga and S atoms were forced to have 4, 4 and 2 neighbors respectively, in line with the results of the unconstrained 'final' model. It was required that at least 95% of the atoms satisfy these constraints.

Different models (see Section 1) could be tested with these simulation runs. M-M bond types were investigated in every possible combination in the case of the $(GeS_2)_{0.75}(Ga_2S_3)_{0.25}$ sample. A model without M-M bonds was also tested. But in this model, gallium atoms were forced to be threefold coordinated. Since in this model only Ge-S and Ga-S bonds are allowed, the fourfold coordinated Ge and twofold coordinated S atoms imply that the average coordination number of Ga must be 3.

The cumulative relative $R$-factors ($R_c$) of different models are compared in Table 4. It was found that regardless the use of coordination constraints, best $R_c$ values were given by models in which Ge-Ge and Ge-Ga bonds are *not* allowed. On the other hand, in case of constrained models the elimination of Ga-Ga bonds resulted in about 20% increase values of $R_c$. Fits of Ga K-edge data with and without Ga-Ga bonds are compared in Fig. 5. From these investigations, it is concluded that Ga-Ga bonds (and Ga-Ga bonds only) are necessary to get reasonable fits.

| allowed M-M bonds | $R_{XRD}$ | $R_{Ge\ EXAFS}$ | $R_{Ga\ EXAFS}$ | $R_c$ |
|---|---|---|---|---|
| Ge-Ge, Ge-Ga, Ga-Ga 'unconstrained' (reference model) | 0.0521 | 0.1429 | 0.0959 | 1 |
| Ga-Ga 'unconstrained' | 0.0348 | 0.1433 | 0.1020 | 0.91 |
| Ge-Ge, Ge-Ga, Ga-Ga | 0.0641 | 0.1315 | 0.1648 | 1.29 |
| Ge-Ge, Ge-Ga | 0.0599 | 0.1098 | 0.2144 | 1.38 |
| Ge-Ge, Ga-Ga | 0.0492 | 0.1156 | 0.1656 | 1.16 |
| Ge-Ga, Ga-Ga | 0.0628 | 0.1356 | 0.1465 | 1.23 |
| Ge-Ge | 0.0519 | 0.1034 | 0.1918 | 1.24 |
| Ge-Ga | 0.0546 | 0.1299 | 0.1688 | 1.24 |
| Ga-Ga | 0.0466 | 0.1353 | 0.1232 | 1.04 |
| None ($N_{Ga}=3$) | 0.0733 | 0.1633 | 0.1100 | 1.23 |

**Table 4** Goodness of the fit values ($R$-factors) of the investigated models. Ge, Ga and S atoms were forced to have 4, 4, and 2 neighbors respectively. The values of the reference model in which all M-M bonds were allowed and coordination constraints were not applied, and 'final' model ('unconstrained' model with Ga-Ga bonds allowed) are also shown for comparison.

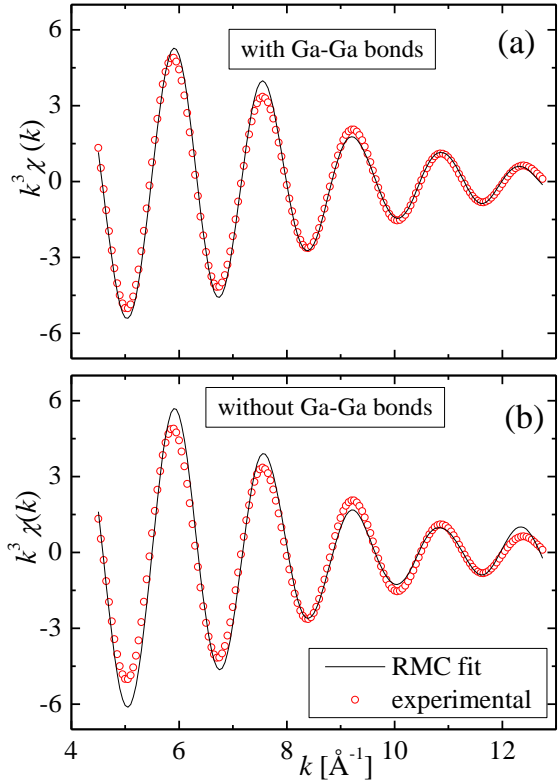

**Figure 5** Ga K-edge EXAFS fits of $(GeS_2)_{0.75}(Ga_2S_3)_{0.25}$ sample with and without Ga-Ga bonds. (a) Ga-Ga bond is allowed, Ge-Ge and Ge-Ga bonds are forbidden; (b) Ge-Ge and Ge-Ga bonds are allowed, Ga-Ga bond is forbidden. In both cases Ge, Ga and S atoms were forced to have 4, 4 and 2 neighbors respectively.

The uncertainty of the average coordination number of gallium was determined in the case of the 'final' model by another series of simulation runs. In these calculations, three coordination constraints were applied: the coordination numbers of Ge and S atoms were forced to be 4 and 2, respectively, while $N_{GaGa}$ was constrained to different values. Since only three types of bonds were allowed (Ge-S, Ga-S, Ga-Ga), the above constraints also determine the total coordination number of gallium. It was found for both compositions that the model curves fit the experimental data properly if $N_{GaGa}=0.7 \pm 0.4$ and resulting total coordination number of Ga is $N_{Ga}=3.7 \pm 0.3$.

The nearest neighbor distances are shown in Table 5 for the different compositions. The Ge-S and Ga-S distances agree with the literary values [40-42]. Due to the lower concentration of Ga, the uncertainty of Ga-Ga bond length is somewhat higher.

|  | $r_{Ge-S}$ (±0.02) | $r_{Ga-S}$ (±0.02) | $r_{Ga-Ga}$ (±0.03) |
|---|---|---|---|
| $(GeS_2)_{0.75}(Ga_2S_3)_{0.25}$ | 2.215 | 2.275 | 2.61 |
| $(GeS_2)_{0.83}(Ga_2S_3)_{0.17}$ | 2.22 | 2.27 | 2.64 |

**Table 5** Nearest neighbor distances of the investigated glasses in Ångström.

The second and third coordination shells around the germanium atoms have been investigated in detail. A series of simulation runs were carried out with the 'final' model and coordination constraints for Ge and S (to have 4 and 2 nearest neighbor, respectively). The simulation box contained 7000 atoms. 30 different starting configurations were tested to reduce statistical errors and estimate inherent uncertainties. The final configurations were analyzed for the number of edge-sharing $[GeS_{4/2}]$ units as well as for the distances between the centers of edge-sharing (ES) and corner-sharing (CS) germanium centered tetrahedra. It was found that 27(-3+6)% and 30(±2)% of Ge atoms can be found in edge-sharing blocks in the $(GeS_2)_{0.75}(Ga_2S_3)_{0.25}$ and $(GeS_2)_{0.83}(Ga_2S_3)_{0.17}$ compositions, respectively (see Table 6).

|  | Ge atoms | S atoms |
|---|---|---|
| $(GeS_2)_{0.83}(Ga_2S_3)_{0.17}$ | 30 (±2) | 16.2 (±1.0) |
| $(GeS_2)_{0.75}(Ga_2S_3)_{0.25}$ | 27 (-3+6) | 14.3 (-1.3+2.1) |

**Table 6** The ratio of Ge and S atoms participating in edge-shared units (in percent). The uncertainties are determined by 30 simulation runs applying different starting configurations.

The relevant Ge-Ge distances are presented in Fig. 6, where their distribution is shown for the corner- and edge-sharing pairs separately. For the corner-sharing Ge-Ge pairs, the relative frequency of this Ge-Ge pair distances has a peak around $r_{Ge-Ge}$= 3.6 Å. This value is a little bit longer than in $GeS_x$ (3.44 Å [59] and 3.5 Å [60]) and $GeS_2$-$Ga_2S_3$ (3.47-3.53 Å [40]) glasses. The distribution of the edge-sharing Ge-Ge distances has a peak around 2.9 Å and a second one at 3.35-3.4 Å, separated by a minimum at 3.1 Å. The first value is similar to what was found in the $Ge_{25}(As,Ga)_{10}S_{65}$ system [41], in crystalline $GeS_2$ [61], and in $GeS_x$ glasses [59, 60]. The third coordination shell around Ge at $r_{Ge-Ge}$ ≈ 3.5 Å found in earlier publications [40, 59, 60] seems to decompose a shorter ES ($r$ = 3.4 Å) and a longer CS ($r$ = 3.6 Å) distances in our simulation models. The split peak of edge-sharing Ge-Ge distances may be caused by differences in the local environment of the participating Ge atoms (e.g. one of the Ge atoms connects to an ethane-like [S$_3$Ga-GaS$_3$] unit).

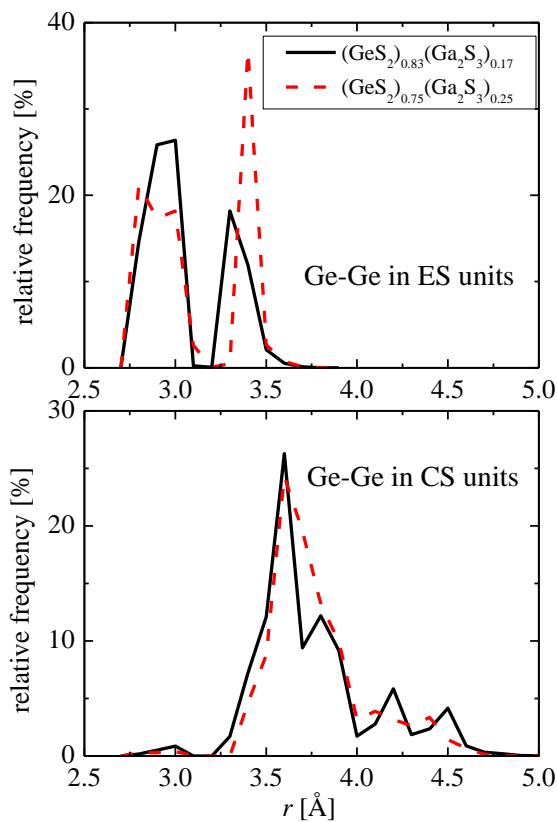

**Figure 6** Relative frequency of pair distances in edge-shared (ES) and corner-shared (CS) GeS$_{4/2}$ units. (Solid line: (GeS$_2$)$_{0.83}$(Ga$_2$S$_3$)$_{0.17}$, dashed line: (GeS$_2$)$_{0.75}$(Ga$_2$S$_3$)$_{0.25}$ composition.)

The effect of the sulfur deficiency is investigated by the connections of the sulfur atoms: if they are participating in edge-share connections or only in corner-share relations. It was suggested that the S-deficiency can be compensated by the formation of edge linkages between [(Ge/Ga)S$_{4/2}$] units [40]. The participation of the sulfur atoms in edge linkages in our samples are presented in Table 6. It was found that the number of sulfur atoms which connect two edge-sharing Ge or Ga based units is not higher in the more sulfur deficient (GeS$_2$)$_{0.75}$(Ga$_2$S$_3$)$_{0.25}$ composition (14%) than in the (GeS$_2$)$_{0.83}$(Ga$_2$S$_3$)$_{0.17}$ sample (16%). This suggests that the higher sulfur deficiency is compensated by the M-M bonding.

### 4.2.3. Comparison with other Ge-Ga-Ch glasses

The total coordination numbers of Ge, Ga and Ch atoms in various Ge-Ga-Ch glasses are collected in

Table 7. Germanium atoms were found to be fourfold coordinated in every Ge-Ga-Ch glass. In the moderately S-deficient Ge-Ga-S glasses, similarly to the Ge-Ga-Se and Ge-Ga-Te glasses, gallium atoms do not obey the 8-N rule; their total coordination number is close to four. While the M-elements behave similarly, the coordination numbers of chalcogens strongly differ in these glasses. In case of sulfur, the coordination number is very close to two, in selenides and tellurides it is around 2.2-2.3.

|  | Ge-Ga-Te | Ge-Ga-Se | Ge-Ga-S |
|---|---|---|---|
| $N_{Ge}$ | 3.8 | 4.0 | 4.0 |
| $N_{Ga}$ | 4.1 | 4.0 | 3.85 |
| $N_{Ch}$ (Ch=Te,Se,S) | 2.25-2.36 | 2.25-2.3 | 2.0 |

**Table 7** Comparison of the total coordination numbers of different Ge-Ga-Ch glasses. Ge-Ga-S from present work, Ge-Ga-Se from [21], and Ge-Ga-Te data from [30]

Besides the coordination number of chalcogens the degree of chemical ordering is also different in Ge-Ga-Ch glasses. Ge-Ga-S glasses can be described by a chemically ordered network model in which Ge-S and Ga-S bonds are preferred. However, as the coordination number of gallium is higher than three the investigated glasses are sulfur deficient, which leads to the appearance of the M-M bonds, for example in [$S_{3/2}$Ga-Ga$S_{3/2}$] ethane-like units (see Figure 7). (It follows from the sample composition that for glasses lying along the GeS$_2$-Ga$_2$S$_3$ tie-line if $N_{GeS}$=4 and $N_{SGe}$+$N_{SGa}$=2 then $N_{GaS}$=3. As N$_{Ga}$=N$_{GaS}$+N$_{GaGa}$=3+N$_{GaGa}$ the excess coordination of Ga and the amount of the Ga-Ga bonds increase together. The presence of the M-M bonds in these samples is a consequence of the chalcogen deficiency.) This is in contrast with Ge-Ga-Se glasses where M-M bonds were reported even in Se-rich compositions, suggesting a less ordered network structure [21]. For example in the slightly Se-rich Ge$_{20}$Ga$_{10}$Se$_{70}$ the Ge-Ga coordination number is as high as ~0.75. As a consequence of the large amount of M-M bonds the Se-Se coordination number is also high (~0.97). (We note that the latter value within the uncertainty agrees with N$_{SeSe}$ in the strongly Se-rich binary Ge$_{20}$Se$_{80}$ glass [22].)

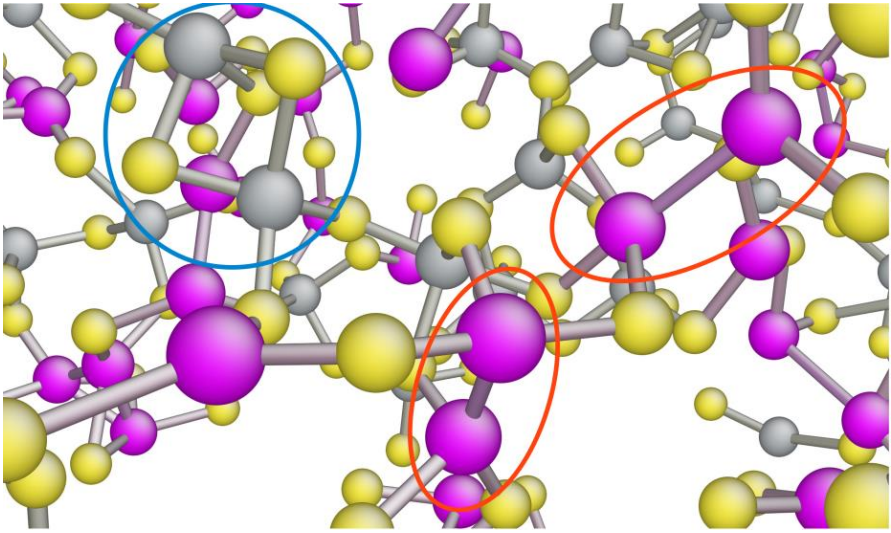

**Figure 7** A part of the 'final' model configuration of the $(GeS_2)_{0.75}(Ga_2S_3)_{0.25}$ composition obtained by RMC simulation. Ge, Ga and S atoms are represented by gray, purple and yellow balls, respectively. (Atomic configuration is visualized by AtomEye program [62].) Some $[S_3Ga-GaS_3]$ and edge-shared $[GeS_{4/2}]$ units are highlighted with red and blue lines.

## 5. Conclusions

$(GeS_2)_{0.75}(Ga_2S_3)_{0.25}$ and $(GeS_2)_{0.83}(Ga_2S_3)_{0.17}$ glasses were investigated by Raman scattering, high energy X-ray diffraction and extended X-ray absorption fine structure measurements at the Ge and Ga K-edges. Reverse Monte Carlo simulation technique was used to obtain large scale structural models by fitting the three experimental data sets simultaneously. It was found that most of the gallium atoms are fourfold coordinated, while germanium and sulfur atoms follow the Mott's rule and have 4 and 2 nearest neighbors, respectively. Sulfur deficiency of these two compositions implies the presence of M-M bonds. The type of these bonds was systematically investigated. It turned out that germanium atoms have only sulfur neighbors and participate in $[GeS_{4/2}]$ structural units. In contrast, Ga-Ga homonuclear bonds are required for a good fitting the experimental data, therefore gallium atoms can be found in $[S_3Ga-GaS_3]$ ethane-like structures as well. Connectivity between $[GeS_{4/2}]$ structural units has been studied and Ge-Ge distances in corner- and edge-shared tetrahedra have been determined. About 30% of germanium atoms were found in edge-shared tetrahedra.

## Acknowledgments

This work was supported by ADEME-French Environment and Energy Management Agency. Authors are thankful to the platform SIR-ScanMAT of University of Rennes 1 for technical assistance in Raman17 spectroscopy. I. Pethes and P. Jóvári were supported by NKFIH (National Research, Development and Innovation Office) Grant No. SNN 116198.